\documentclass[sigconf]{acmart}

\AtBeginDocument{%
  \providecommand\BibTeX{{%
    \normalfont B\kern-0.5em{\scshape i\kern-0.25em b}\kern-0.8em\TeX}}}

\usepackage{color} 

\copyrightyear{2022}
\acmYear{2022}
\setcopyright{acmcopyright}\acmConference[CHI '22 Extended Abstracts]{CHI Conference on Human Factors in Computing Systems Extended Abstracts}{April 29-May 5, 2022}{New Orleans, LA, USA}
\acmBooktitle{CHI Conference on Human Factors in Computing Systems Extended Abstracts (CHI '22 Extended Abstracts), April 29-May 5, 2022, New Orleans, LA, USA}
\acmPrice{15.00}
\acmDOI{10.1145/3491101.3519712}
\acmISBN{978-1-4503-9156-6/22/04}

\title{\toolName{}: Interactive Analysis and Validation of Suspected Rumors on Social Media}
\newcommand{\toolName}{\textit{RumorLens}}
\newcommand{\name}{{\textit{RumorLens}}}
\author{Ran Wang}
\affiliation{%
  \institution{School of Journalism and Information Communication, Huazhong University of Science and Technology}
  \country{China}
}

\author{Kehan Du}
\affiliation{%
  \institution{School of Computer Science, United Kingdom University of Bristol}
  \streetaddress{1 Th{\o}rv{\"a}ld Circle}
  \country{United Kingdom}}

\author{Qianhe Chen}
\affiliation{%
  \institution{School of Computer Science and Technology, Huazhong University of Science and Technology}
  \country{China}}

\author{Yifei Zhao}
\affiliation{%
  \institution{School of Artificial Intelligence and Automation, Huazhong University of Science and Technology}
  \country{China}
}

\author{Mojie Tang}
\affiliation{%
 \institution{School of Architecture and Urban Planning, Huazhong University of Science and Technology}
 \country{China}}

\author{Hongxi Tao}
\affiliation{%
  \institution{Computer Science Department, University College London}
  \country{United Kingdom}}
  
\author{Shipan Wang}
\affiliation{%
  \institution{School of Journalism and Information Communication, Huazhong University of Science and Technology}
  \country{China}
  \postcode{78229}}

\author{Yiyao Li}
\affiliation{%
  \institution{School of Journalism and Information Communication, Huazhong University of Science and Technology}
  \country{China}}

\author{Yong Wang}
\affiliation{%
  \institution{School of Computing and Information Systems, Singapore Management University}
  \country{Singapore}}

\renewcommand{\shortauthors}{Ran Wang et al.}

\begin{document}
\begin{abstract}

With the development of social media, various rumors can be easily spread on the Internet and such rumors can have serious negative effects on society. Thus, it has become a critical task for social media platforms to deal with suspected rumors. However, due to the lack of effective tools, it is often difficult for platform administrators to analyze and validate rumors from a large volume of information on a social media platform efficiently. We have worked closely with social media platform administrators for four months to summarize their requirements of identifying and analyzing rumors, and further proposed an interactive visual analytics system, \toolName{}, to help them deal with the rumor efficiently and gain an in-depth understanding of the patterns of rumor spreading. \toolName{} integrates natural language processing (NLP) and other data processing techniques with visualization techniques to facilitate interactive analysis and validation of suspected rumors. We propose well-coordinated visualizations to provide users with three levels of details of suspected rumors: an overview displays both spatial distribution and temporal evolution of suspected rumors; a projection view leverages a metaphor-based glyph to represent each suspected rumor and further enable users to gain a quick understanding of their overall characteristics and similarity with each other; a propagation view visualizes the dynamic spreading details of a suspected rumor with a novel circular visualization design, and facilitates interactive analysis and validation of rumors in a compact manner. By using a real-world dataset collected from Sina Weibo, 
one case study with a domain expert is conducted to evaluate \toolName{}. The results demonstrated the usefulness and effectiveness of our approach.
\end{abstract}





\maketitle
\begin{figure*}[htbp]
  \includegraphics[width=\textwidth]{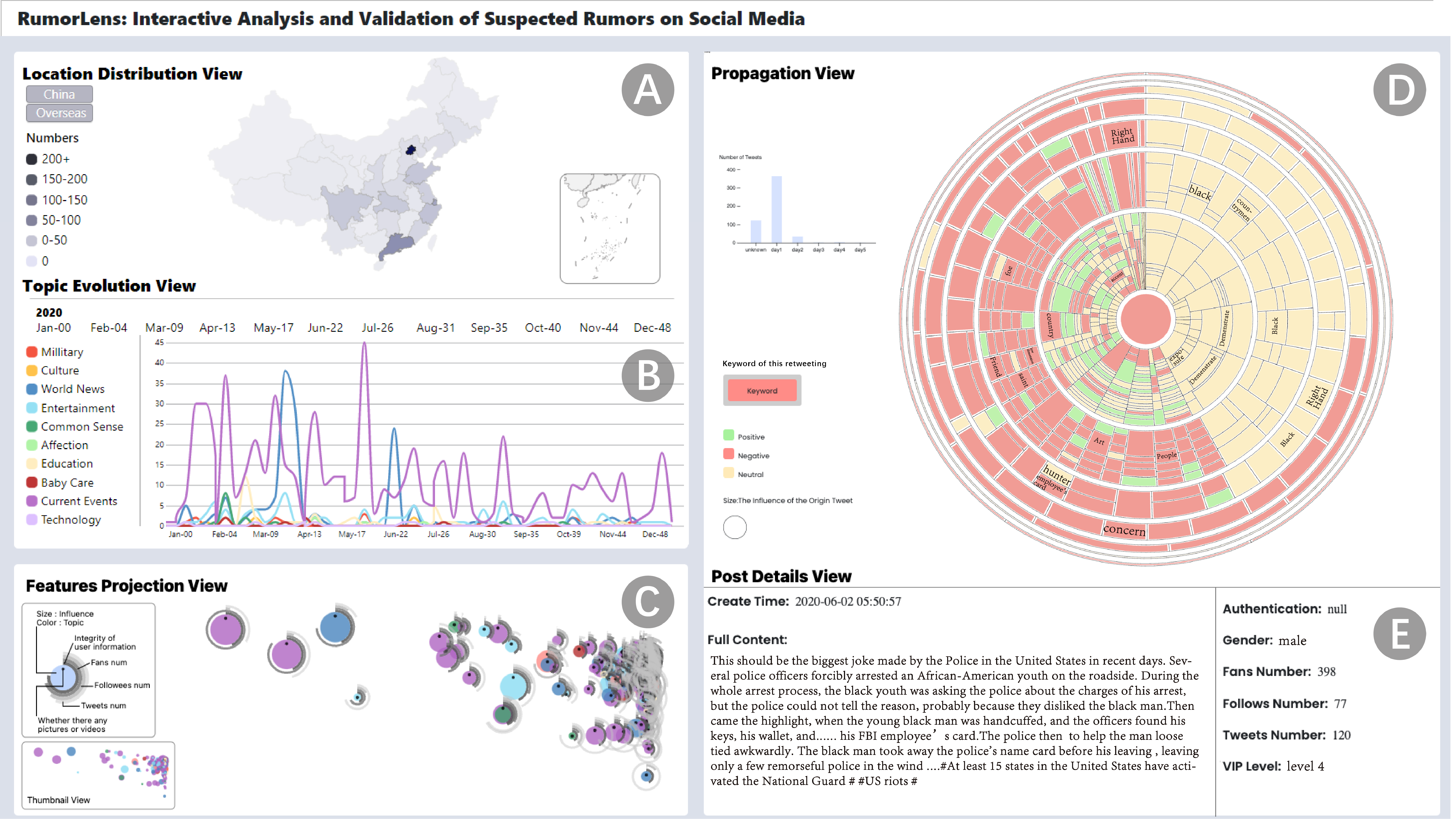}
  \caption{\toolName{}, a multi-level visual analytics system to help users analyze and validate suspected rumors on social media in an interactive way. A) Location Distribution View provides a summary of suspected rumors' spatial distribution; B) Topic Evolution View shows changes of suspected rumors with different topics over time; C) Features Projection View reveals overall characteristics of suspected rumors and similarity with each other; D) Propagation View visualizes the dynamic spreading details of a suspected rumor with a novel circular visualization design; E) Post Details View displays the details of user information and post contents.}
  \Description[The overview of our RumorLens]{Figure 1. Fully described in the text.}
  \label{fig:toutu}
 \end{figure*}

\section{Introduction}
Social media has been widely used in our everyday life and made information sharing and communication very convenient. However, it also provides an easy and fast way to generate and spread various rumors~\cite{shu2017fake,shelke2019Source}.
Rumors often refer to the information that
has already been
spread among people, but has not been verified or is intentionally spread to mislead the audience~\cite{allport1947psychology, qazvinian2011rumor, zhao2015enquiring, ma2018rumor}.
Such rumors on social media can have serious negative effects on both individuals~\cite{difonzo2007Rumor,del2016spreading} and the society~\cite{ma2019visual,boididou2017Detection,allcott2017social}, for example, causing public panic, disrupting the social order, decreasing the government credibility, and even endangering national security~\cite{Gang2015Rumor}.
Service providers of social media have tried to identify the rumor spread on social media platforms
in the past few years~\cite{dang2019early}.

Conventional approaches for rumor identification rely on the personal experience of content specialists. Though high accuracy can be guaranteed by this kind of manual approach, it is quite time-consuming and labor-intensive to deal with
massive of information. 
In addition, there have been more and more automatic rumor identification methods~\cite{dayani2015rumor,ma2017detect,li2019rumor}, where rumor identification is defined as a binary classification problem. Such automatic rumor identification methods are often more efficient than manual approaches.
However, most existing automatic methods can not ensure accuracy and are often not robust in practical applications~\cite{2019A}. Compared with those automatic methods, the experience of human beings can provide a richer output that includes the reason for the decision~\cite{2019A}.
Currently, the approaches for rumor identification based on human-computer collaboration are more preferred~\cite{Gang2015Rumor}.
For example, computer algorithms~\cite{2014TweetCred} or crowdsourcing techniques are first employed to detect suspected rumors on social media, which will be validated by a team of journalists in a manual way subsequently~\cite{zubiaga2018detection}.

These approaches enable platform administrators of social media to improve their efficiency for rumor identification.
However, after several rounds of in-depth interviews with domain experts from social media,
we identified several major issues and concerns of using those rumor identification approaches.
Above all, suspected rumors have been identified by crowdsourcing or automatic detector at first~\cite{2019A}.
Nevertheless, due to the lack of effective tools, it is still a challenge for social media platform administrators to deal with a large number of suspected rumors, especially for those regarding emergencies with considerable impact.
Second, it is necessary for platform administrators to gain an in-depth understanding of suspected rumor features in terms of content, user, topic, and propagation, which makes rumor validation solid and convincing. In particular, it is 
important to trace the dynamic propagation details of a suspected rumor on social media~\cite{wu2015false}, which, however, is difficult to achieve now.
In this study, we aim to develop novel visual analytic techniques to address these challenges.

To this end, we propose an interactive visual analytic system, \name{}, to help the administrators of social media platforms to analyze and validate suspected rumors efficiently and gain deep insights into the rumor spreading.
In particular, we develop well-coordinated visualization views to provide users with details of the suspected rumors at different levels: 
(1) an overview of suspected rumors is displayed for filtering suspected rumors rapidly, which visualizes the spatial distribution and temporal evolution of rumors; 
(2) a projection view is designed for quick selection based on their overall features and similarity with each other, and each suspected rumor is represented with a metaphor-based glyph;
(3) a propagation view is provided to analyze and validate suspected rumors interactively, where we propose a novel circular visualization design to visualize the dynamic spreading details of the selected suspected rumor in a compact manner.
Furthermore, \name{} enables linked exploration and smooth interactions across the three major views so that our target users can explore
the suspected rumors from multiple perspectives and different levels of detail.
To extensively evaluate \name{}, we conducted a case study to demonstrate the usefulness and effectiveness of a dataset collected from Sina Weibo, a popular social media in China.
The main contributions of this paper are summarized as follows:
\begin{itemize}
    \item An interactive visualization system, \name{}, to assist social media platform administrators in analyzing and validating suspected rumors efficiently with multiple levels of details.
    \item A novel circular design to show the dynamic spreading details of a suspected rumor, facilitating interactive analysis and validation of rumors in a compact manner.
    \item A case study conducted by a domain expert to demonstrate the usefulness and effectiveness of \name{} in helping users gain deep insights into suspected rumors on social media.
\end{itemize}

\section{Related Work}

Our related work can be categorized into three parts:
rumor detection in social media, visual analytics of social media, and visualization of information diffusion.


\textbf{Rumor Detection in Social Media:} 
Depending on whether the human is involved in the process of rumor identification, rumor detection can generally be classified into two types: manual methods and automatic methods. 
Manual methods include (1) expert-based method and (2) crowd-sourced based method\cite{zhang2020overview}. 
For the expert-based method, experts who have a deep knowledge of certain domains will manually annotate tweets as `true', `false', or `unverified'~\cite{ahsan2019rumors}. They validate rumors with the help of fact-checking websites~\cite{dang2019early} and inquiry retweets~\cite{zhao2015enquiring}.
For crowd-sourced fact-checking, any user on Sina Weibo can report suspicious rumors, and the final declaration would be made by a group of experts~\cite{Gang2015Rumor}.
However, manual methods only rely on human judgment and efforts. Thus, they cannot handle a large volume of data~\cite{zubiaga2018detection}. To address the scalability, automatic rumor detection techniques have been developed, leveraging natural language processing and machine learning techniques~\cite{dayani2015rumor,ma2018rumor}. Features are first extracted and then identified by various kinds of classifiers~\cite{2016Fake,ito2015assessment,zhao2015enquiring,2017Simple,2016Learning,vosoughi2015automatic}. The key issue focus on integrating multiple rumor features and are committed to selecting the most effective features~\cite{dang2019early,castillo2011information}
which are 
divided into four types~\cite{castillo2011information}: (1) content features~\cite{li2019rumor}; 
(2) user features, such as fans number, followees number~\cite{dayani2015rumor}; (3) topic features, which are aggregates computed from content and user features~\cite{sahana2015automatic}; and (4) propagation features, which consider attributes related to the propagation path~\cite{del2016spreading,ma2016detecting}.
However, none of them can deal with all rumors with different topics and dynamic propagation effectively. Meanwhile, manual labeling is a very challenging and time-consuming task.
In this paper, we propose a visual analyzing approach that combines human domain knowledge and automated machine learning techniques to make a rumor analysis and validation reliable and explainable.

\textbf{Visual Analytics of Social Media:}
Prior studies in visual analytics of social media most focus on information collection and behaviors understanding~\cite{wu2016survey}.
In the field of information collection, it can be generally summarized into three categories: keyword-based~\cite{archambault2011themecrowds,wang2018towards}, topic-based~\cite{Bernstein2010Eddi,zhao2014fluxflow} and multi-faceted approaches~\cite{wu2014opinionflow}. Research studies about behaviors understanding are proposed to help understand user's behaviors, including collective 
behaviors~\cite{viegas2013google+,fu2018visforum}, cooperation and competition~\cite{FB2004,2014EvoRiver}. These approaches enable a multi-aspect analysis of a large collection of tweets through data visualization and rich interactions, such as filtering, zooming, and clustering, etc.
However, few studies have been conducted to visually analyze and validate suspected rumors on social media in the tight cooperation of theoretical research and experience experts from social media.

\textbf{Visualizations of Information Diffusion:} \label{Visualizations of Information Diffusion Patterns}
Two main types of data visualization techniques have been proposed to represent information diffusion on social media, i.e., network-based (e.g., node-link diagram) and space-filling-based (e.g., Voronoi, treemap) approaches~\cite{wu2016survey}. For network-based methods, nodes and links are employed to visualize a social network, allowing users to understand opinion flow or explore propagation patterns~\cite{heer2005vizster,cao2012whisper,ma2019visual}. However, 
such network-based approaches suffer from scalability issues
when the network on social media scales to tens of thousands of nodes.
To address this issue, space-filling-based methods have been proposed. Adjacency diagrams encode connectivity between nodes with an adjacency matrix rather than individually drawn edges~\cite{2006Balancing,2009SmallBlue}. Nonetheless, the time series and hierarchical relationship among tweets are usually ignored, which are quite important for rumor validation. Inspired by such visualizations, we propose a novel circular design to show the dynamic spreading of a suspected rumor in a compact manner.

\section{Data Abstraction and Processing}
\label{dataprocessing}


Sina community management center is an online platform in charge of collecting, recording and dealing with illegal messages including suspected rumors published on Sina Weibo, one of the most famous social media in China. As reported by Sina Weibo, tens of thousands of rumors in various fields have to be dealt with in time every year, or else they may result in a serious impact on both the public and the government~\cite{wu2015false}.
We collected a dataset of suspected rumors between 2019/12/27 and 2020/12/14 from the center, with 936 suspected rumors, approximately 80000 corresponding retweets and comments, and 53843 user profiles.
Followed by suggestions of platform administrators and previous studies mentioned above, multiple features regarding user, content, topic and propagation of suspected rumors are important for rumor analysis and validation. Hence, in addition to the original data of tweets, TF-IDF~\cite{ramos2003using}, sentiment recognition~\cite{8684825}, topic classification~\cite{devlin2018bert}, influence calculation~\cite{bakshy2011everyone} are employed to further extract keywords, sentiment, topic and influence of suspected rumors respectively. In particular, we adopt t-SNE~\cite{van2008visualizing} to reduce the feature dimension of suspected rumors and project onto a 2D map to show relationships among them.

\section{Requirements Analysis} \label{RA}

To better understand the major challenges and design requirements for suspected rumor analysis and validation, we have worked closely with three platform administrators (E1-E3) from social media for four months. They were experts from three famous internet companies (Sohu, NetEase, and TouTiao) respectively. E1 is in charge of rumor collection and validation in Sohu for two years. E2 is responsible for content production in NetEase and familiar with features of rumors. E3 is a platform administrator who is mainly checking and publishing content in TouTiao. All the experts have experience in dealing with rumors on the Internet. By conducting a series of interviews and discussions with them, we collected their feedback and summarized the major design requirements \textbf{R1-R3} as follows.

\textbf{R1. Explore the Overall Space-Time Distribution of Suspected Rumors.} Based on the feedback of our experts, all of them agreed that it is almost impossible to manually review all the messages within a limited time. According to their work principles, those recent suspected rumors concerning people's livelihood are supposed to be dealt with with higher priorities. Therefore, an effective approach should help easily and quickly filter suspected rumors with priorities, especially those could have negative effects. E1 suggested that spatial-temporal attributes and the topic of the message should be considered.

\textbf{R2. Inspect Suspected Rumor Cases Through Characteristics Comparisons.} All the experts agreed that the characteristics analysis of rumors can provide significant clues for rumor validation. E2 pointed out that the messages whose characteristics are different from others are more likely rumors, especially for those ones with significant influence. In addition, according to E3's work experience, pictures and videos are usually added for rumors to enhance their credibility. Thus, visual inspection of suspected rumors in feature space can help platform administrators select a case with the most suspicious.

\textbf{R3. Explore the Propagation Details of Individual Suspected Rumors.} As E3 suggested, a functionality of exploring the propagation details of a suspected rumor is essential. It can help platform administrators further confirm suspected rumor cases, which may be wrongly labeled by some automated detection methods or reported by users from social media. E1 and E2 agreed that propagation path, contents of tweets and retweets are important factors for rumor validation. For example, rumors with negative sentiment diffused significantly faster and deeper than the truth. Also, some users from social media are eager to refute a rumor in comments. For such kind of exploration, a convenient and compact way to validate a suspected rumor is highly appreciated. 

\section{Visual Design}

As mentioned above, we also proposed straightforward and intuitive visual designs to help platform administrators analyze and validate suspected rumors based on their distribution, features and propagation. 

\subsection{Suspected Rumors Overview}
Suspected Rumors Overview (Figure~\ref{fig:toutu}(A) and (B)) is designed to provide platform administrators with an overview of all suspected rumors in terms of spatial-temporal distribution \textbf{(R1)}. The Suspected Rumors Overview is composed of two main parts, a choropleth map and a line chart. The choropleth map (Figure~\ref{fig:toutu}(A)) displays numbers of suspected rumors across different parts of China and overseas as a whole. The line chart (Figure~\ref{fig:toutu}(B)) displays the number of changes of suspected rumors with different topics over time. 

In the choropleth map on the top, the location with more suspected rumors is shown in a darker color and the exact number of which can be observed by hovering on it. When clicking on the map, corresponding locations are highlighted and selected. The line chart below shows the topic evolution of suspected rumors selected on the choropleth map, with different colored lines representing different topics. Once a topic is chosen, the keywords of suspected rumors under the topic are shown upon the corresponding line. A time axis is also provided to filter suspected rumors further by choosing a start and end time.

\subsection{Projection View}

Projection View (Figure~\ref{fig:toutu}(C)) is designed to help platform administrators quickly inspect and locate the most suspicious features, which may belong to a rumor, for further validation \textbf{(R2)}. The Projection View consists of glyphs representing filtered suspected rumors projected onto a 2D feature map, where the distance between glyphs indicates the similarity between suspected rumor messages.

\begin{figure}[htbp]
    \centering
    \includegraphics[width=0.5\textwidth]{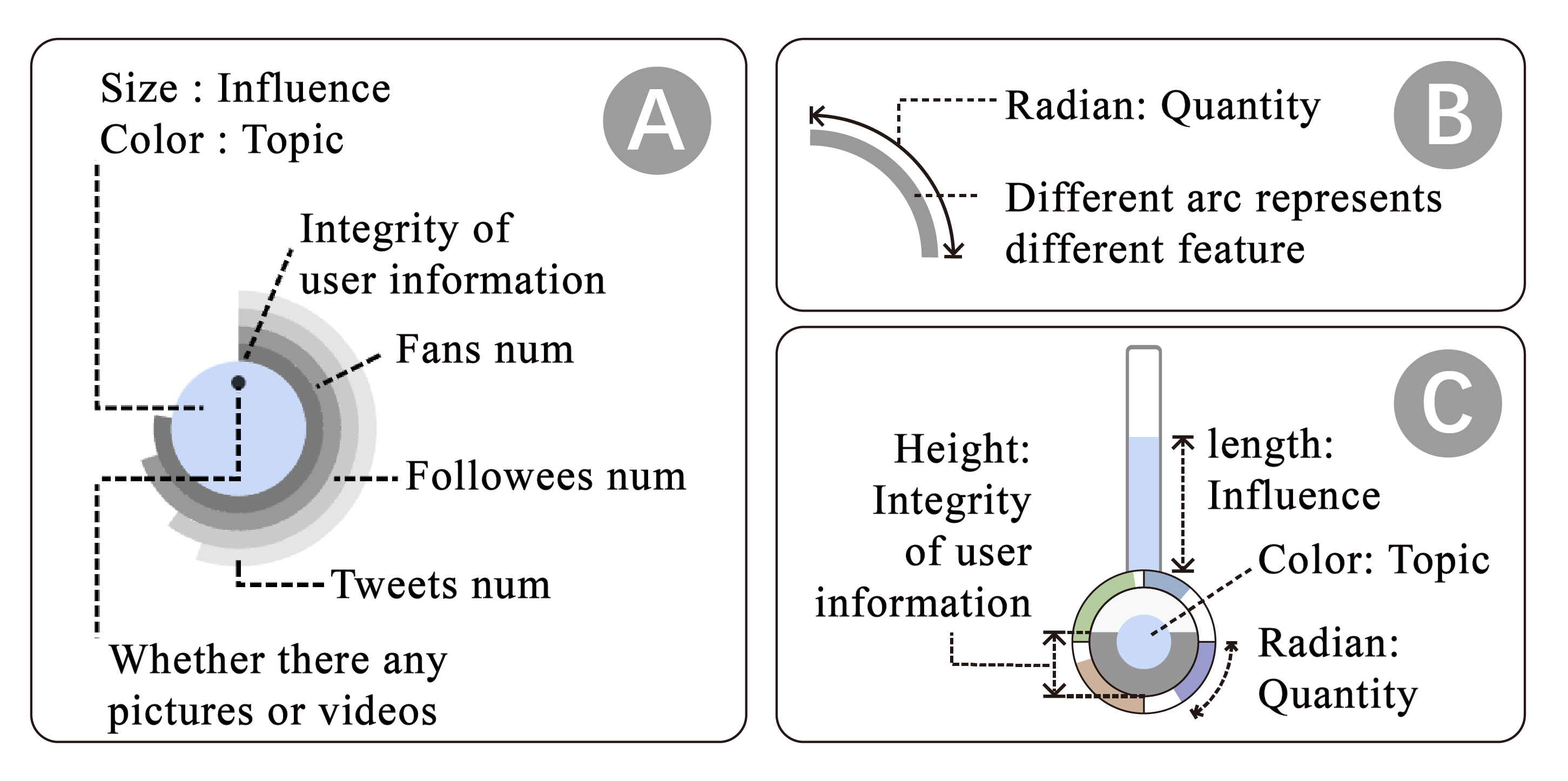}
    \caption{Glyph designs for features of each suspected rumor. (a) round glyph design; (b) arc glyph design; (c) thermograph shape glyph design.}
    \Description[Glyph design comparison for features projection]{Figure 2. Fully described in the text.}
    \label{fig:projection_glyph}
\end{figure}

For facilitating visual inspection and comparison of suspicious features, we encode each suspected rumor as a round glyph which is composed of two parts: the inner circle and the outer sectors, as shown in Figure~\ref{fig:projection_glyph} (A). The color of the inner circle represents the topic of the suspected rumor, while the size indicates its influence. The outer four arcs reveal the number of fans, followees, tweets, and integrity of user information respectively, and the glyph is illustrated in Figure~\ref{fig:projection_glyph} (B). In particular, since the number of fans and followees varies greatly among different users, they are calculated by the logarithmic approach for convenient comparison. Compared with the thermograph shape (Figure~\ref{fig:projection_glyph} (C)), our glyph design can provide a more concise and compact way to show each suspected rumor. Besides, under the guidance of thumbnails on the left side, the position and size of the view displayed can be adjusted by dragging and zooming. 
Once selected, the glyph will be magnified to highlight the differences from others for close observation.

\subsection{Propagation View}
Propagation View (Figure~\ref{fig:toutu} (D) and (E)) provides a detailed understanding of how a suspected rumor message diffuses on social media, which enables platform administrators to make a final decision \textbf{(R3)}. It is of great importance since rich information in propagation mentioned by experts' experience and previous studies can be displayed and explored for suspected rumor validation. The view contains two parts, a novel circular design to visualize the suspected rumor propagation at the top and a table showing corresponding content details at the bottom. 

\begin{figure*}[htbp]
    \centering
    \includegraphics[width=0.62\textwidth, height = 3.5cm]{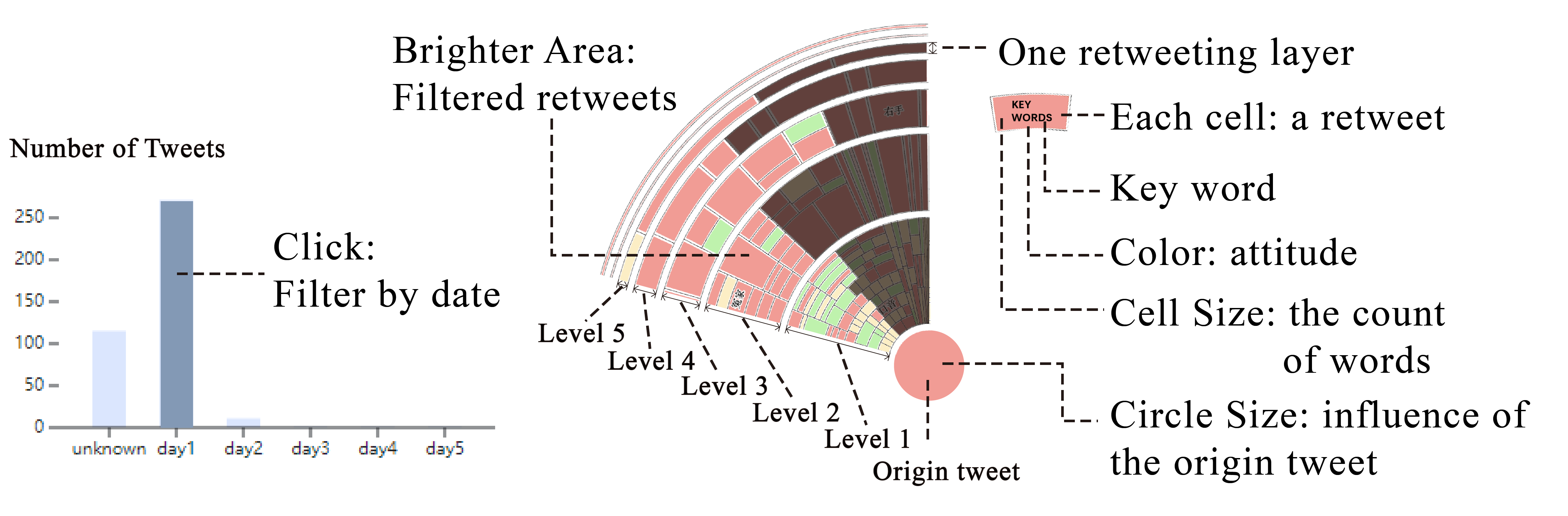}
    \caption{A glyph design for Propagation View. The graph is used to show the specific encoding and interaction demonstration.}
    \Description[Glyph encoding of propagation of a suspected rumor]{Both propagation hierarchy and time line can be visualized in a compact manner, and retweets can be filtered and highlighted by the time histogram on the left.}
    \label{fig:propagation1}
\end{figure*}

We propose a novel circular design and it consists of the original tweet in the center and its retweets of multi-levels in the surrounding. As shown in Figure~\ref{fig:propagation1}, the original tweet represented by a circle node is located in the center, where size indicates its influence. Around it, there are several concentric circular rings with different widths, each denoting the retweets at the same hierarchy. Every concentric circular ring contains hands of sectors separated by radial white lines, which are arranged clockwise by dates. Each sector is fully filled with retweets encoded by a number of cells, the size of which indicates the number of words contained. Also, the cell's color represents whether the sentiment of a retweet is neutral (yellowish), positive (green) or negative (red). In particular, keywords able to provide rough meanings of the retweets are shown within some large cells. 

\begin{figure*}[htbp]
    \centering
    \includegraphics[width=0.70\textwidth, height = 7.0cm]{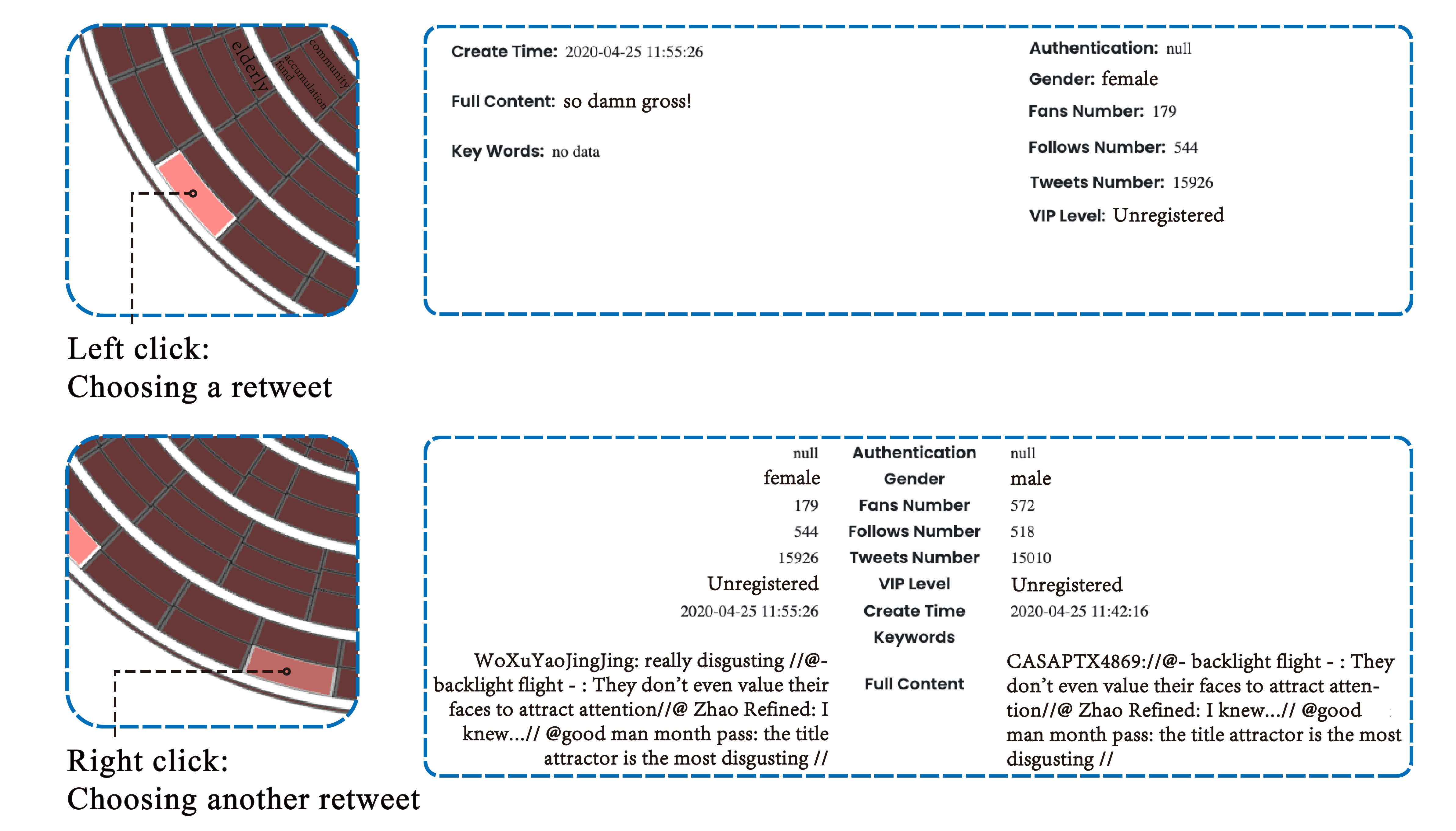}
    \caption{Interaction Design of the Propagation View.
    It allows users to select two retweets through left and right clicks.
    The Detail View will show the detailed information of the selected two retweets for a detailed comparison.}
    \Description[Interactions on the propagation view]{Multiple retweeting units are allowed to choose for details comparison.}
    \label{fig:inter1}
\end{figure*}

Rich interactions are supported in Propagation View. First, a panel would appear to display some brief user information when hovering on the corresponding cell, and more detailed information can be displayed in Figure~\ref{fig:toutu}(E) through clicking, such as user attributes, create time and full content. Second, a comparative analysis is also supported to facilitate a careful inspection. By clicking two cells of retweets successively, information in detail of both can be viewed and checked side by side, as shown in Figure~\ref{fig:inter1}. Third, there is a time histogram at the top left corner showing changes in the retweeting number over time. Platform administrators can click on it and highlight cells retweeted on the same day at different hierarchies.

During our design process, we initially considered using a node-link diagram, tree-map, spiral timeline or sunburst chart to visualize a tweet propagation on social media. However, none of them can present retweeting hierarchies and time series from the propagation path simultaneously. Besides, a node-link diagram is not space-efficient enough when a considerable amount of data needs visualization. In comparison, our proposed circular design is able to show the dynamic spreading details of a suspected rumor, facilitating interactive analysis and validation of rumors in a compact manner.

\section{Case Study}

We conducted a case study on a suspected rumor dataset of Sina Weibo to demonstrate the effectiveness of our proposed visual design. In this case study, E1, who has rich experience in dealing with rumors, was invited to find a suspected rumor of his interest and further validate whether it is actually a rumor or not.


\textbf{Overall Exploration.} 
At first, E1 loaded all the suspected rumors into \toolName{}, and started to interact with the choropleth map in Figure~\ref{fig:toutu}(A). He first clicked the button of \textit{overseas} and filtered all the suspected rumors from overseas into the topic evolution view shown in Figure~\ref{fig:toutu}(B). As the number of suspected rumors with different topics was changed and displayed over time, E1 paid attention to a blue curve with a relatively high peak, which belonged to the topic of \textit{World News}. After checking through the keywords along the curve, he clicked the label representing \textit{World News} on the left for further exploration. 

\textbf{Feature Inspection.} 
Then he moved to the Projection View to inspect the suspected rumors of \textit{World News} on a feature map. Because the glyphs are too crowded to choose, as shown in Figure~\ref{fig:toutu}(C), E1 dragged and zoomed in for better observation until a comparatively bigger glyph than its surroundings is found. According to figure legends on the left, this suspected rumor was of significant influence, yet with an unauthenticated user account and a relatively small number of fans. Since the features mentioned above highly match his work experience, E1 selected the suspected rumor for further validation in the propagation view.

\textbf{Instance Validation.}
In Figure~\ref{fig:toutu}(D), he found that there were 6 white circular rings, demonstrating that the depth of retweets reached as many as 6, which was deeper than most normal circumstances. Moreover, most of the colors on the view are red and green, which means the retweets showed strong sentiment inclinations, especially for the negative. Then he figured out what happened by clicking the node at the center, as shown in Figure~\ref{fig:toutu}(E) and checked keywords floating on some cells for more detail. Most of them were related to street demonstrations, indicating that some social security incidents may have happened yet not focused on the incident itself directly. He continued to explore until a keyword called ID card was found, which may be connected tightly to the incident. When clicking on the cell, a table with detailed information showed the full list of comments saying that the original tweet was misleading. Several retweets supported the viewpoint.

Finally, E1 checked the details of the original tweet again for further validation. It was mainly about an American African who was accidentally arrested and then turned out to be an FBI detective. He found the content of the tweet contains some apparent ambiguous words like \textit{approximately}, which was used to excuse the activity of police officers. According to the social support theory~\cite{zhou2021linguistic},
rumors with persuasive and uncertain words are more likely to be disseminated on social media. Individuals with similar experiences are inclined to obtain social support and start to trust those kinds of rumormongers and even nurtured to retweet the rumors. At last, After taking into consideration all the above factors, E1 confidently concluded that the tweet is probably a rumor and should be further dealt with.


\section{Conclusion and Future Work}

We have proposed \name{}, an interactive visual analytics system to help administrators of social media platforms deal with suspected rumors efficiently. To tackle the challenges of validating suspected rumors, a novel visual interactive system is proposed to provide deep insights into multiple levels of details of the suspected rumors. In addition, a novel circular glyph design is proposed to show the dynamic spreading details of a suspected rumor, facilitating interactive analysis and validation of rumors in a compact manner. 
A case study conducted by a domain expert provides support for the effectiveness of \name{} in helping users analyze and validate the suspected rumors. 

However, \name{} targeting on interactive analysis and validation of suspected rumors still needs further improvement. First, by working with domain experts, we realize the importance of user information for rumor identification. For example, the message has a high risk of being a rumor if the account of the user is undefined and there are identified rumors tweeted several times before. Hence, it is necessary to provide more information on historical complaints related to the user. Second, rumors can be identified by various kinds of features, and thus how to choose and evaluate their effects on rumor validation is still a problem to be solved.


\name{} is currently designed for and evaluated through the dataset collected from Sina Weibo, where the language is Chinese.
In future work, we would like to extend \name{} to other languages and further evaluate its effectiveness with datasets of other languages.
Also, our approach is mainly focusing on the text information of rumor messages, where the image information of some rumor messages is ignored. It will be interesting to further incorporate the image information in the analysis and validation of suspected rumors on social media.
\balance
 \begin{acks}
 We thank all the domain experts for their valuable feedback, and the anonymous reviewers for their constructive comments. This research is founded by National Social Science Fund of China award number(s): 19CXW032.
 The computation is completed in the HPC platform of Huazhong University of Science and Technology. 
 Yong Wang is the corresponding author (yongwang@smu.edu.sg).  
\end{acks}


   
   

\bibliographystyle{ACM-Reference-Format}


\end{document}